\documentclass[11pt]{article}
\usepackage{graphicx}

\usepackage{geometry}
\usepackage{setspace}
\usepackage{graphicx}
\usepackage{amssymb}
\usepackage{epstopdf}
\usepackage{multirow}
\usepackage{extarrows}
\usepackage{rotating}
\usepackage{dsfont}
\usepackage{bm}
\usepackage{mathabx}
\usepackage{endnotes}
\usepackage{tikz}
\newcommand\mathC{\mkern1mu\raise2.2pt\hbox{$\scriptscriptstyle|$}
        {\mkern-7mu\rm C}}              

\newtheorem{definition}{Definition}[section]
\newtheorem{theorem}{Theorem}[section]

\newtheorem{proposition}[theorem]{Proposition}
\newtheorem{corollary}[theorem]{Corollary}

\newcommand{\disjoint}
  {\ensuremath{\mathop{
    \begin{tikzpicture}[line width=0.12ex]
      \useasboundingbox (-1ex, -1ex) rectangle (1ex, 1ex);
      \draw (-1ex,0.4ex) -- (0,0.4ex) -- (0,-1ex) -- (1ex,-1ex);
    \end{tikzpicture}}
  }\nolimits}

\title{Is mereology empirical? Composition for fermions} 
	             
\begin{document}
\maketitle

\begin{abstract}
How best to think about quantum systems under permutation invariance is a question that has received a great deal of attention in the literature. But very little attention has been paid to taking seriously the proposal that permutation invariance reflects a representational redundancy in the formalism.  Under such a proposal, it is far from obvious how a constituent quantum system is represented.  Consequently, it is also far from obvious how quantum systems compose to form assemblies, i.e.~what is the formal structure of their relations of parthood, overlap and fusion.

In this paper, I explore one proposal for the case of fermions and their assemblies.  According to this proposal, fermionic assemblies which are not entangled---in some heterodox, but natural sense of `entangled'---provide a \emph{prima facie} counterexample to classical mereology.  This result is puzzling; but, I argue, no more intolerable than any other available interpretative option.
\end{abstract}

\tableofcontents

\section{Introduction}

The fact that quantum mechanics forces a revision of many of our dearly held metaphysical beliefs is by now familiar.  In this article, I aim to provide one more example of such a metaphysical belief; namely that classical mereology---by which I mean the formal theory of parts and wholes developed by, amongst others, Leswenski (1916), Tarski (1929) and Leonard \& Goodman (1940)---gives a true account of the structure of composition for physical objects. 

One might suppose that the culprit responsible for the failure of mereology in the quantum domain is entanglement.  As is well known, any entangled state fails to supervene on (i.e.~be determined by) the states of the joint system's constituents---supposing that we happy attributing the constituents with states at all.\footnote{I am picking up here on the fact that, in any entangled state, the constituents' states, as calculated by performing a partial trace on the joint state, are improper, i.e.~not ignorance-interpretable, mixtures.}  This has prompted some (e.g. Maudlin (1998)) to claim that a certain strong version of reductionism fails for quantum systems.

Could this failure of reductionism lead to a failure of mereology? Surely not.  For the failure of reductionism is merely a failure of the joint system's properties (as encapsulated by its state) to supervene on the constituents' \emph{properties} (as encapsulated by \emph{their} states).  And the possibility that there may be irreducible relations was countenanced at least as far back as Russell (1918).  Thus, entanglement merely demands that we accept irreducible relations into our ontology of quantum mechanics. (A point also argued by Ladyman \& Ross (2007, 149-50).)

The problem posed for mereology which I wish here to present arises specifically for fermions, which include the particles which ``make up'' all stable matter---in \emph{some} sense of ``make up''!  The argument relies on adopting a heterodox, but nevertheless correct, interpretation to an important constraint in quantum mechanics, known as \emph{permutation invariance}.  According to this interpretation, permutation invariance reflects a representational redundancy in the standard formalism of quantum mechanics, somewhat analogous to the representational redundancy  that counterpart theorists see in standard Tarskian semantics, in which two distinct models can differ purely as to which object plays which role in the pattern of instantiation of properties and relations.

This interpretation of permutation-invariant quantum mechanics prompts a revision to some common quantum concepts, most importantly for us here, what counts as an entangled state.  On this revision, some   joint states of fermions count as not entangled.  The contradiction with mereology will be proven for these states alone:  thus  the failure of mereology  has nothing do with entanglement.

The argument will run as follows.  Upon interpreting the permutation invariance of fermionic joint states in the correct way, we shall see that these states may be represented naturally by subspaces of the single-system Hilbert space.  More specifically, a non-entangled joint state of $N$ fermions will correspond to an $N$-dimensional subspace.  A natural definition of parthood will emerge, which is then represented by the relation of subspacehood.  The associated notion of mereological fusion between fermionic assemblies---which can be defined using parthood alone---will then be shown \emph{not} to match the structure that smaller fermionic assemblies (i.e.~assemblies containing fewer fermions) bear to larger ones (i.e.~assemblies containing more fermions).

Some strategies to save mereology in the face of this result will be considered.  As I will argue, one successful strategy exists, in the sense that the truth of the axioms of mereology may be preserved.  However,  it will remain the case that the operation which unifies many fermions into one fermionic assembly is \emph{not} mereological fusion.

The title of this article is intended to reference the famous discussion of quantum logic, inspired by the work of Birkhoff and von Neumann (1936),  in Putnam (1968, 1974) and  Dummett (1976), and continued by Maudlin (2005) and Baciagaluppi (2009).  Indeed we shall see that the analogy here is exact: fermionic composition is to quantum logic as mereological composition is to classical logic.  That is, while it is well-known that the mathematical structures used to describe the relationship between objects in mereology or propositions  in classical logic are the same (namely, Boolean algebras), so too the  the mathematical structures used to describe the relationship between \emph{fermions} or propositions  in \emph{quantum} logic are the same (namely, Hilbert lattices).

The structure of the article is as follows.   In Section \ref{PIQM}, I discuss briefly the issue of permutation invariance in quantum mechanics.  In Section \ref{Trans}, I outline classical mereology and offer a means to ``translate'' the states of quantum mechanics into Tarskian models, so that the question of whether mereology holds of them can meaningfully be put.  Section \ref{Results} contains the main results, and considers one  saving strategy for mereology.

\section{Permutation-invariant quantum mechanics}\label{PIQM}

In this Section I will introduce the key formal and interpretative basics for our discussion of fermionic composition.  The main motivation is to define (or redefine) \emph{entanglement}
in a permutation-invariant setting in a physically salient way, particularly for fermions.  I will argue that, under the most appropriate definitions of these terms, there are fermionic states that are not entangled; these states will be the focus of the results of Section \ref{Results}. 

\subsection{Permutation invariance and its interpretation}\label{PI}

Permutation-invariant quantum mechanics is standard quantum mechanics with the additional condition of permutation invariance.  We begin with the single-system Hilbert space $\mathcal{H}$.  From this we define the $N$-fold tensor product $\otimes^N \mathcal{H}$, the \emph{prima facie} state space for $N$ ``indistinguishable'' systems (their indistinguishability is expressed by the fact that any two factor Hilbert spaces are unitarily equivalent).  The joint Hilbert space $\otimes^N \mathcal{H}$ carries a natural unitary representation $U:S_N\to\mathcal{U}(\otimes^N \mathcal{H})$ of the group $S_N$ of permutations on $N$ symbols.  For example, the permutation $(ij)$, which swaps systems $i$ and $j$, is represented by the unitary operator $U(ij)$ defined on basis states (having chosen an orthonormal basis $\{|\phi_k\rangle\}$ on $\mathcal{H}$) by
\begin{eqnarray}
&& U(ij)|\phi_{k_1}\rangle\otimes\ldots\otimes|\phi_{k_i}\rangle\otimes\ldots\otimes|\phi_{k_j}\rangle\otimes\ldots\otimes|\phi_{k_N}\rangle
\nonumber\\
&& \qquad\qquad\qquad\qquad\qquad =
|\phi_{k_1}\rangle\otimes\ldots\otimes|\phi_{k_j}\rangle\otimes\ldots\otimes|\phi_{k_i}\rangle\otimes\ldots\otimes|\phi_{k_N}\rangle
\end{eqnarray}
and then extended by linearity.  Permutation invariance, otherwise known as the \emph{Indistinguishability Postulate} (Messiah \& Greenberg 1964, French \& Krause 2006), is the condition  on any operator $Q\in\mathcal{B}(\otimes^N \mathcal{H})$, that it be \emph{symmetric};\footnote{This use of `symmetric' is not to be confused with the condition that $\langle\psi|Q\phi\rangle = \langle Q\psi|\phi\rangle$ for all $|\psi\rangle, |\phi\rangle\in\mbox{dom}(Q)$.} i.e.~for all permutations $\pi\in S_N$ and all states $|\psi\rangle\in\otimes^N \mathcal{H})$,
\begin{equation}
\langle\psi|U^\dag(\pi)QU(\pi)|\psi\rangle = \langle\psi|Q|\psi\rangle
\end{equation}
The representation $U$ is reducible, and decomposes into several copies of inequivalent irreducible representations, each irreducible representation corresponding to a different \emph{symmetry type}; namely bosonic states, fermionic states and (if $N\geqslant 3$) a variety of paraparticle states (see e.g.~Tung 1985, Ch.~5). If we consider only the information provided by the symmetric  operators, we treat permutation invariance as a superselection rule, and each  superselection sector corresponds to one of these symmetry types.  

What does it mean to ``impose'' permutation invariance?  Isn't it rather that permutation invariance holds of some operators and not others?  I propose that imposing permutation invariance means to lay it down as a necessary condition on any operator's receiving a physical interpretation.  This  justifies, and is justified by, treating the factor Hilbert space labels---i.e.~the order in which single-system operators and states lie in the tensor product---as nothing but an artefact of the mathematical formalism of quantum mechanics.  

What is the justification for interpreting the factor Hilbert space labels in this way?  Of course, the ultimate justification  is that it leads to an empirically adequate theory.   It is an empirical fact that elementary particles exhibit statistics consistent with their being either bosons or fermions.  But this fact is logically weaker than the claim that factor Hilbert space labels represent nothing.  It \emph{could} be that factor Hilbert space labels represent (or name), for example, the constituent systems, and that the joint state of any assembly of elementary particles remains in the fermionic or bosonic sector under all actual physical evolutions due only to the fact that the corresponding Hamiltonian happens to be permutation-invariant.   Indeed, this interpretative gloss is offered by many authors (e.g.~French \& Redhead 1988; Butterfield 1993; Huggett 1999, 2003; French \& Krause 2006; Muller \& Saunders 2008; Muller \& Seevnick 2009; Caulton 2013).  However, it may be argued that the physical emptiness of the factor Hilbert space labels offers the \emph{best explanation} of the empirical fact that permutation invariance seems always to hold true.  This suggestion is in line with a more general interpretative stance in physics: that any exact symmetry is a symptom of representational redundancy in the corresponding theory's formalism.

\subsection{Fermionic states and GMW-entanglement}\label{GMW}

The focus of this paper is fermionic states and their compositional structure.  Picking some orthonormal basis $\{|\phi_i\rangle\}$ in $\mathcal{H}$, these states are spanned by states of the form
\begin{equation}
\frac{1}{\sqrt{N!}}\sum_{\pi\in S_N} (-1)^{\deg\pi}|\phi_{i_{\pi(1)}}\rangle\otimes|\phi_{i_{\pi(2)}}\rangle\otimes\ldots \otimes|\phi_{i_{\pi(N)}}\rangle
\end{equation}
and carry the alternating irreducible representation of $S_N$; i.e.~any permutation $\pi$ is represented by multiplication by $(-1)^{\deg\pi}$, where $\deg\pi$ is the \emph{degree} of the permutation $\pi$ (i.e.~the number of pairwise swaps into which  $\pi$ may be decomposed).

 Following  Ladyman, Linnebo and Bigaj (2013), we may use the mathematical apparatus of \emph{Grassmann} or \emph{exterior algebras} to represent fermionic states.  The exterior algebra $\Lambda(V)$ over the vector space $V$ (over the field of complex numbers $\mathbb{C}$)  is obtained by quotienting the tensor algebra $T(V) := \bigoplus_{k=0}^{\infty} T^k(V) = \mathbb{C}\oplus V \oplus (V\otimes V) \oplus (V\otimes V\otimes V) \oplus \ldots$ with the equivalence relation  $\sim$ defined so that $\alpha\sim\beta$ iff $\alpha$ and $\beta$ have the same anti-symmetrization;\footnote{Equivalently, $\Lambda(V)$ is the quotient algebra $T(V)/D(V^2)$ of $T(V)$ by the two-sided ideal $D(V^2)$ generated by all 2-vectors of the form $x\otimes x$. See e.g.~Mac Lane \& Birkoff (1991, \S XVI.6).} i.e.
\begin{equation}
    \Lambda(V) := T(V)/\sim\ .
\end{equation}
For example, $[x\otimes y] = [-y\otimes x]$ and $[x \otimes x] = [\mathbf{0}]$.
We may set $V=\mathcal{H}$, then there is  a natural isomorphism  $\iota$ from the elements of $\Lambda(\mathcal{H})$ onto the vectors of the fermionic Fock space $\mathcal{F_-(H)} := \bigoplus_{N=0}^{\dim\mathcal{H}} \mathcal{A}(\otimes^N\mathcal{H})$. $\iota$ simply takes any $\sim$-equivalence class of degree-$r$ vectors of $T^r(\mathcal{H})$ to the anti-symmetric degree-$r$ vector in $\mathcal{A}(\otimes^r\mathcal{H})$ that is their common anti-symmetrization.  Therefore we may pick out any $N$-fermion state in $\mathcal{A}(\otimes^N\mathcal{H})$ by specifying its pre-image under $\iota$ in $\Lambda^N(\mathcal{H})$ (i.e.~the subalgebra of $\Lambda(\mathcal{H})$ containing only degree-$N$ vectors).

Elements of $\Lambda(V)$ are called \emph{decomposable} iff they are equivalence classes $[x_{i_1}\otimes x_{i_2}\otimes \ldots\otimes x_{i_r}]$ containing product vectors. Not all elements are decomposable; an example is given in Section \ref{}.
 The product on the exterior algebra is the \emph{exterior} or \emph{wedge product} $\wedge$, defined by its action on decomposable elements as follows: 
  \begin{equation}\label{wedgeprod}
[x_{i_1}\otimes x_{i_2}\otimes \ldots\otimes x_{i_r}] \wedge [x_{i_{r+1}}\otimes x_{i_{r+2}}\otimes \ldots\otimes x_{i_{r+s}}] = [x_{i_1}\otimes x_{i_2}\otimes \ldots\otimes x_{i_{r+s}}] \ ,
 \end{equation}
where  $\{x_1,x_2, \ldots x_{\dim V}\}$ is an orthonormal basis for $V$ and each $i_k\in \{1,2,\ldots, \dim V\}$.  We then extend the definition of $\wedge$ to non-decomposable elements by bilinearity.  (Note that if there is a pair ${i_j} = i_k$ for $j\neq k$, then the righthand side of (\ref{wedgeprod}) is $[\mathbf{0}]$.) For any $\alpha\in\Lambda^r(V)$ and any $\beta\in\Lambda^s(V)$,
 $
 \alpha \wedge \beta = (-1)^{rs}\beta\wedge\alpha \in \Lambda^{r+s}(V)
$.

In the following, I will, like Ladyman, Linnebo and Bigaj (2013), make use of a harmless abuse of notation by referring to anti-symmetric states by their corresponding wedge product.  In particular, given an orthonormal basis $\{|\phi_i\rangle\}$ on $\mathcal{H}$, 
\begin{equation}
|\phi_{i_1}\rangle\wedge|\phi_{i_2}\rangle\wedge\ldots \wedge|\phi_{i_N}\rangle
\end{equation}
will be used as a shorthand for
\begin{equation}
\frac{1}{\sqrt{N!}}\sum_{\pi\in S_N}(-1)^{\deg\pi}|\phi_{i_{\pi(1)}}\rangle\otimes|\phi_{i_{\pi(2)}}\rangle\otimes\ldots \otimes|\phi_{i_{\pi(N)}}\rangle\ .
\end{equation}
The distinction between decomposable and non-decomposable fermionic states has a clear analogy with the distinction between product and non-product states.  However, decomposable fermionic states have the property, unlike product states, that there are (up to a possible factor of $-1$) invariant under arbitrary permutations in their factor space indices.  Therefore the wedge product offers a permutation-invariant way of constructing joint states of, say $N$ fermions, out of $N$ fermion states, much as the tensor product offers a permutation \emph{non-invariant} way of constructing joint states for ``distinguishable'' systems.

The analogies between the tensor product in the permutation-non-invariant case and the wedge product in the fermionic case suggest  a redefinition of entanglement for fermionic states.  The standard definition, which which we  do not take issue in the permutation-non-invariant case, is that an assembly's state is entangled iff it is non-separable, i.e.~it cannot be written as a product state (see e.g.~Nielsen \& Chuang 2010, 96).   This suggests redefining entanglement for fermions so that \emph{a fermionic joint state is entangled iff it is not decomposable}, in the sense given above.

In fact this redefinition has been suggested already, by Ghirardi, Marinatto and Weber in a series of papers (Ghirardi, Marinatto \& Weber 2002, Ghirardi \& Marinatto 2003, 2004, 2005), and endorsed by Ladyman, Linnebo \& Bigaj (2013).\footnote{The definition that Ghirardi, Marinatto \& Weber actually offer is equivalent to the one above.}  Therefore I  call the proposed notion \emph{GMW-entanglement}.  Further discussion of the physical salience of this notion is taken up in Caulton (2015).

The important fact for Section \ref{Results} is that decomposable fermionic states have a feature that is not shared by non-entangled states under the standard definition.  That is that decomposable fermionic states, corresponding as they do to decomposable elements of the exterior algebra on $\mathcal{H}$, correspond to \emph{subspaces} of $\mathcal{H}$.  More specifically the state
\begin{equation}
|\phi_{1}\rangle\wedge|\phi_{2}\rangle\wedge\ldots \wedge|\phi_{N}\rangle\ ,
\end{equation}
where $\langle\phi_i|\phi_j\rangle = \delta_{ij}$, corresponds to the $N$-dimensional subspace spanned by the degree-1 vectors $|\phi_1\rangle, |\phi_2\rangle, \ldots |\phi_N\rangle$.  This offers a glimpse of  two of our main results in Section \ref{}, namely: (i)  parthood between fermionic assemblies is represented by subspacehood; and (ii) the state of a larger assembly is given by the \emph{span} of the states of its constituents.  The tension  between the two notions of fusion implicit in (i) and (ii) embodies the tension between classical mereology and the quantum mechanics of fermions.

\section{Setting up quantum mechanics for mereology}\label{Trans}

\subsection{Classical mereology}

There are several axiomatizations of classical mereology available (see Hovda (2009) and Varzi (2014) for a discussion); for the purposes of this paper, I have  chosen the one that most allows the most perspicuous discussion of its troubles for fermionic systems.
Classical mereology requires only one primitive term, $\sqsubseteq$ (parthood).  From this we  define \emph{proper parthood}:
\begin{equation}
\forall x\forall y(x\sqsubset y \leftrightarrow (x\sqsubseteq y\ \&\ x\neq y)) ,
\end{equation}
the \emph{overlap} relation $x\circ y$ (`$x$ overlaps $y$') in terms of common parthood:
\begin{equation}
\forall x\forall y(x\circ y \leftrightarrow \exists z(z\sqsubseteq x\ \&\ z\sqsubseteq y))
\end{equation}
and the \emph{disjointness} relation $x\disjoint y$ as the contrary of overlap:
\begin{equation}
\forall x\forall y(x\disjoint y \leftrightarrow \neg x\circ y)
\end{equation} 
Finally, given any 1-place formula $\phi$, something is a \emph{fusion} of the $\phi$s iff all and only its overlappers overlap some $\phi$.  So we define $\mathcal{F}_\phi(x)$ (`$x$ is a fusion of the $\phi$s') as follows:
\begin{equation}
\forall x(\mathcal{F}_\phi(x) \leftrightarrow \forall y(y\circ x \leftrightarrow \exists z(\phi(z)\ \&\ z \circ y)))
\end{equation}
With these definitions, we may now present the two axioms and one axiom schema.  I also include a third axiom, \emph{Axiomicity}, which is not essential to classical mereology, but which will hold in all of the theories discussed here.
\begin{enumerate}
\item \emph{Partial Order}.  $\sqsubseteq$ is a partial order (i.e.~it is reflexive, anti-symmetric and transitive).

\item \emph{Strong Supplementation}. If something is not a part of a second thing, then some part of the first thing is disjoint from the second thing:
\begin{equation}
\forall x\forall y(x\not\sqsubseteq y \rightarrow \exists z(z\sqsubseteq x\ \&\  z \disjoint y))
\end{equation}

\item \emph{Atomicity}. Everything has a part that has no proper parts.
\begin{equation}
\forall x\exists y(y\sqsubseteq x\ \&\ \neg\exists z z \sqsubset y)
\end{equation}

\item \emph{Unrestricted Fusion}. If there are some $\phi$s, then there is a fusion of the $\phi$s:
\begin{equation}
(\exists x\phi(x) \rightarrow \exists x\mathcal{F}_\phi(x))
\end{equation}
This is imposed for all substitution instances of $\phi$.
\end{enumerate}

\subsection{Finding the subsystems in the quantum formalism}\label{Subs}
It will be key to proving the results in Section \ref{Results} that we have some way of identifying in the quantum formalism when the joint system has subsystems in particular states.  This requires giving some physical interpretation to that formalism.  In this we are constrained by the requirements of permutation-invariance to give a physical interpretation \emph{only} to those quantities which are permutation invariant.

We assume that we are dealing with an $N$-fermion assembly, so the joint state lies in $\mathcal{A}\left(\otimes^N\mathcal{H}\right)$, where $\mathcal{H}$ is the single-system Hilbert space.   We expect any subsystem's state to lie in $\mathcal{A}\left(\otimes^r\mathcal{H}\right)$, where $1\leqslant r\leqslant N$. I will categorise projectors according to the Hilbert space they act on.  A projector is of degree-$r$ iff it acts on $\mathcal{A}\left(\otimes^r\mathcal{H}\right)$ (where $r=1$ corresponds to the single-system Hilbert space $\mathcal{H}$).

Choose any degree-1 projector $P$.    Its orthocomplement is $P_\perp := \mathds{1}-P$.  From $P$ we may  define  a family of projectors $\{\sigma^s_r(P)\ |\ 1\leqslant r\leqslant s\leqslant N \}$:
\begin{eqnarray}
\sigma^s_0(P) &:=& \underbrace{P_\perp \otimes\ldots\otimes P_\perp}_{s}\nonumber\\
\sigma^s_1(P) &:=& P\otimes \underbrace{P_\perp \otimes\ldots\otimes P_\perp}_{s-1}
\ +\
  P_\perp\otimes P \otimes\underbrace{\ldots\otimes P_\perp}_{s-2}
 \ +\
 \ldots
 \ +\
 \underbrace{P_\perp \otimes\ldots\otimes P_\perp}_{s-1} \otimes P \nonumber\\
 \sigma^s_2(P) &:=& P\otimes  P \otimes \underbrace{P_\perp \otimes\ldots\otimes P_\perp}_{s-2}
\ +\
  P\otimes P_\perp\otimes P \otimes\underbrace{\ldots\otimes P_\perp}_{s-3}\nonumber\\
  &&\qquad\qquad\qquad\qquad\qquad\qquad\qquad\qquad
 \ +\
 \ldots
 \ +\
 \underbrace{P_\perp \otimes\ldots\otimes P_\perp}_{s-2} \otimes P \otimes P\nonumber\\
 & \vdots \nonumber\\
 \sigma^s_s(P) &:=&  \underbrace{P \otimes\ldots\otimes P}_{s}
\end{eqnarray}
These projectors will be the most important ones below.
Each $\sigma^s_r(P)$ is a symmetric projector, and so may be interpreted as corresponding to a physical property.  I propose the following interpretation:
\\
\\
\indent  $\sigma^s_r(P)$ corresponds to the property `Exactly $r$ of $s$ degree-1 constituents have property $P$'.
  \\\\
 This interpretation can obviously be justified in the context in which permutation invariance is \emph{not} imposed. In that case, each summand of $\sigma^s_r(P)$, which acts on exactly $r$ of $s$ factor Hilbert spaces with $P$ and on the remaining $s-r$ with $P_\perp$, can itself be given a physical interpretation, according to which some selection of $r$ named degree-1 systems have property $P$ and the remaining degree-1 systems do not.  The sum over all summands can then be interpreted as a (quantum) disjunction over all possible selections of $r$ named degree-1 systems.
 
 However, when permutation invariance \emph{is} imposed, this justificatory story is not available to us.  For the individual summands of $\sigma^s_r(P)$ are typically not themselves permutation-invariant, and so, as per our discussion in Section \ref{PIQM}, cannot receive a physical interpretation.  Instead, the physical interpretation offered above must be taken as primitive. 
  
   It is  worth pointing out some formal properties of $\sigma^s_r(P)$, which are consistent with the interpretation offered.  First, it must be emphasised that the domain of  $\sigma^s_r(P)$ is restricted to $\mathcal{A}\left(\otimes^s\mathcal{H}\right)$: so, in particular, if $d:=$ dim$(P) \geqslant s$, then dim$\left(\sigma^s_s(P)\right) = \left(\begin{array}{c}d\\s\end{array}\right)$; otherwise $\sigma^s_s(P) = 0$, due to Pauli exclusion.  Second, we have that  $\sigma^s_r(P) = \sigma^s_{s-r}(P_\perp) $, so  exactly $r$ of $s$ degree-1 constituents have the property $P$ iff exactly $s-r$ degree-1 constituents have the property $P_\perp$, which is the quantum negation of $P$.  Third,
due to Pauli exclusion, $\sigma^s_r(P)=0$ if dim$(P) <r$, or, since $\sigma^s_r(P) = \sigma^s_{s-r}(P_\perp) $, if dim$(P_\perp) = d - \mbox{dim}(P) < s-r$, where $d:=$ dim$(\mathcal{H})$. So a non-vanishing $\sigma^s_r(P)$ requires $r\leqslant $ dim$(P)\leqslant d-s+r$. 

An important result for later will be
\begin{proposition}\label{ProjResult}
For any  degree-1 projectors $P,Q$:  $P\leqslant Q$ iff $\sigma^s_s(Q)\leqslant \sigma^s_r(P)$, where $r := \dim(P)$ and $s:=\dim(Q)$.  
\end{proposition}
 \emph{Proof.} \\
 (\emph{Left to Right}.) Since $\sum_{i=0}^s\sigma^s_i(P) = \sigma^s_s(\mathds{1})$, which is the identity on $\mathcal{A}\left(\otimes^s\mathcal{H}\right)$, we can multiply $\sigma^s_s(Q)$ on the right with the identity to obtain
\begin{equation}\label{blah}
\sigma^s_s(Q) =  \sigma^s_s(Q)\left(\sum_{i=0}^s\sigma^s_i(P)\right) =  \sum_{i=0}^s\sigma^s_s(Q)\sigma^s_i(P) .
\end{equation}
dim$(P) = r$, so $\sigma^s_i(P) = 0$ for $i>r$; so at most the first $r$ terms of this sum are non-vanishing.   Now decompose $Q$ into $Q = P + R$, where $R:=P_\perp Q = QP_\perp$.  Then $\sigma^s_s(Q)\sigma^s_i(P) = \sigma^s_s(Q)\sigma^s_{s-i}(P_\perp) = \sigma^s_s(Q)\sigma^s_{s-i}(R)$. But dim$(R) = s-r$, so $\sigma^s_{s-i}(R) = 0$ for $s-i>s-r$, i.e., $i<r$; so at most the last $s-r$ terms of the sum in (\ref{blah}) are non-vanishing.  It follows that the only non-vanishing term in (\ref{blah}) is for $i=r$; so $\sigma^s_s(Q) =  \sigma^s_s(Q)\sigma^s_r(P)$.  By multiplying $\sigma^s_s(Q)$ on the left with the identity, we can similarly show that $\sigma^s_s(Q) =  \sigma^s_r(P)\sigma^s_s(Q)$.  It follows that $\sigma^s_s(Q)\leqslant \sigma^s_r(P)$.

\noindent (\emph{Right to Left.})
$\sigma^s_s(Q)\leqslant \sigma^s_r(P)$ means that $\sigma^s_r(P)\sigma^s_s(Q) = \sigma^s_s(Q)\sigma^s_r(P) = \sigma^s_s(Q)$.  It follows that $P$ and $Q$ commute. Define $S:=QP = PQ$. Assume for \emph{reductio} that $P\nleqslant Q$; then, since dim$(P) = r$, it must be that dim$(S)<r$.  In that case $\sigma^s_r(S) = 0$, due to Pauli exclusion.  But $\sigma^s_s(Q)\sigma^s_r(S) = \sigma^s_s(Q)\sigma^s_r(P) = \sigma^s_s(Q)$ and dim$(\sigma^s_s(Q)) = 1$; so dim$(\sigma^s_r(S))\geqslant 1$.  Contradiction. So we must have $P\leqslant Q$.  $\Box$

The physical interpretation of this result is as follows: if all $s$ constituents satisfy $Q$, then exactly $r$ of $s$ satisfy $P$.  We can understand this as a result of Pauli exclusion.

\subsection{Translation rules}

The question whether mereology holds true or not for quantum mechanics is \emph{prima facie} ill-formed: mereology is a theory axiomatised in a first-order formal language, while quantum mechanics has no first-order axiomatisation and is instead presented in the mathematics of linear operators on Hilbert spaces.  Therefore we require some way to ``translate'' the claims of one theory into the framework of the other.  It will be simplest to run the direction of translation from quantum mechanics to mereology.

More specifically, we will set up a correspondence between the states of fermionic assemblies, which are normalised vectors in some Hilbert space, and Tarskian models.  This correspondence will be constrained by what I call ``translation rules''.  I hasten to add that the goal is \emph{not} to do quantum mechanics in first-order logic!  The goal is simply to represent the states of quantum mechanics in a form appropriate for comparison with mereology.

First some general remarks regarding the objects and properties of the ``translated'' quantum states:

\emph{Objects.}
The domain of any model will contain only two kinds of objects: quantum systems and projectors.  In the following, I will use lower case variables to range over quantum systems and upper case variables to range over projectors.  (This is just a notational convenience: our models are  first-order, and both kinds of object are objects in the Frege-Quine sense of belonging to the first-order domain.)  Any model will have the total system in its domain.

\emph{Predicates.}
There will only be two primitive predicates.  The first is $\sqsubseteq$, denoting the mereological parthood relation, already discussed.  For our purposes, we may stipulate that this relation  holds (if at all) only between quantum systems.
The second primitive predicate is $E(x, P)$, which denotes a certain dyadic relation between a quantum system $x$ and projector $P$.  We stipulate that $E$ never holds between two quantum systems (e.g.~$E(x,y)$) or two projectors (e.g.~$E(P,Q)$), or between a projector and a system in the wrong order (e.g.~$E(P,x)$).  $E(x,P)$ has the following  interpretation: \emph{$x$ has the property associated with $P$.} 

The translation rules are now presented as follows:
\begin{enumerate}
\item (\emph{Total System}). 

 The total system $\Omega$, whose state is represented by a ray in $\mathcal{A}(\otimes^N\mathcal{H})$, exists.
 
\item (\emph{Existence and Completeness of Projectors}).

 All and only \emph{symmetric} projectors of rank $r$, where $r\in\{1,2,\ldots, N\}$, exist.

\item (\emph{Eigenstate-Eigenvalue Link}).

For any system $x$ and any projector $P$: $E(x, P)$ iff $x$'s state is an eigenstate of $P$ with eigenvalue 1.

\item (\emph{Existence of Subsystems}).

For any degree-1 projector $P$ and all $r = 1,2,\ldots, N$:  $E\left(\Omega, \sum^N_{i=r}\sigma^N_i(P)\right)$ iff there is some system $x$ such that $E(x, \sigma^r_r(P))$.

\item (\emph{Uniqueness of Subsystems}).

For any degree-1 projector $P$ and all $r = 1,2,\ldots, N$:  $E\left(\Omega, \sigma^N_r(P)\right)$ iff there is some unique system $x$ such that $E(x, \sigma^r_r(P))$.

 \item (\emph{Non-GMW-Entangled Systems}). 

For all systems $x$, there is some $r\in\{1,2,\ldots, N\}$ and some degree-1 projector $P$ with dim$(P) = r$ such that $E(x,\sigma^r_r(P))$.

\item (\emph{Definition of Parthood}).

For any quantum systems $x$ and $y$, $x\sqsubseteq y$ iff: for any degree-1 projector $P$ and  all $s\in\{1,2,\ldots, N\}$, if  $E(y, \sigma^s_s(P))$, then  there is some $r\leqslant s$ such that $E(x, \sigma^r_r(P))$.

\end{enumerate}
Some discussion about these rules is in order.  I take each one in turn.
\begin{enumerate}
\item (\emph{Total System}). 

 This rule ensures that the total system $\Omega$ belongs to the domain.
 
\item (\emph{Existence and Completeness of Projectors}).

This rule expresses two essential interpretative assumptions.  The first is that the quantum formalism is \emph{complete}, so that no physical facts are left unrepresented by the quantum state.  The second assumption is none other than the interpretation of permutation invariance as underpinned by representational redundancy, as discussed in Section \ref{PIQM}.

\item (\emph{Eigenstate-Eigenvalue Link}).

This rule also expresses the completeness of the quantum formalism.  It has a controversial element, which is that it applies not only to the total system $\Omega$, but also all subsystems; see below.

\item (\emph{Existence of Subsystems}).

This is the only rule which introduces systems other than $\Omega$ into the domain.  The guiding idea, following from the discussion in Section \ref{Subs}, is that \emph{at least} $r$ of the total system's $N$ degree-1 constituents have some property $P$ iff there is at least one system all  $r$ of whose degree-1 constituents have property $P$.   However, we will see later that this interpretation cannot quite be correct, if by `constituent' we mean atomic part.

\item (\emph{Uniqueness of Subsystems}).

The guiding idea here is that exactly $r$ of $N$ degree-1 constituents of the total system have property $P$ iff  there is a unique system all $r$ of whose degree-1 constituents have property $P$. The existence-entailing component is redundant, given (\emph{Existence of Subsystems}), but is included here for expedience.

This rule permits us to extend (\emph{Eigenstate-Eigenvalue Link}) to subsystems, in the following way.  Given a unique $x$ such that $E(x, \sigma^r_r(P))$, we may say that $x$ has a states whose corresponding density operator has its domain and range in the range of $\sigma^r_r(P)$ (which is a projector).  For any (degree-$r$) projector $Q$ such that $Q\sigma^r_r(P)Q = \sigma^r_r(P)$, we may infer  $E(x, Q)$. 

 \item (\emph{Non-GMW-Entangled Systems}). 

Since dim$(P) = r$, dim$(\sigma^r_r(P)) = 1$.  In fact the range of $\sigma^r_r(P)$ is the ray spanned by the degree-$r$ non-GMW-entangled state $|\phi_1\rangle\wedge|\phi_2\rangle\wedge\ldots\wedge|\phi_r\rangle$, where  $\{|\phi_i\rangle\}$ is any family of orthonormal degree-1 states which span the range of $P$.  So this rule entails that \emph{all} systems occupy decomposable, i.e.~non-GMW-entangled, states.

This rule is problematic if $N\geqslant 4$.  For, in that case, it is not true that any non-GMW-entangled state can be decomposed only into states that are themselves decomposable.
  This corresponds to the well known result for exterior algebras that the non-decomposable degree-2 vector $\xi := \frac{1}{\sqrt{2}}(a\wedge b + c\wedge d)$ satisfies $\xi\wedge \xi = a\wedge b\wedge c\wedge d$.
However, we may take this rule as a \emph{restriction of the domain} to those systems which \emph{are} non-GMW-entangled.  All future reference to systems is then to be taken as implicitly concerning only non-GMW-entangled systems.

\item (\emph{Definition of Parthood}).

This connecting principle can only be justified for non-GMW-entangled fermionic systems.  The idea  is that $x$ is a part of $y$ iff all the degree-1 constituents of $x$ are also constituents of $y$, so if all of $y$'s degree-1 constituents have some property $P$, then \emph{a fortiori} all of $x$'s degree-1 constituents have that same property.
\end{enumerate}

\section{Composition for fermions}\label{Results}

We are now in  a position to establish the main results of this paper.  They are presented in Section \ref{MResults}.  Section \ref{Saving} contains a concluding discussion.

\subsection{Main results}\label{MResults}

\begin{proposition}[Unique Degree]\label{RankU}
Every system has a unique degree in $\{1,2,\ldots N\}$; i.e.~if $E(x,P)$ and $E(x,Q)$, then $\deg(P) = \deg(Q)$. 
\end{proposition}
 \emph{Proof.}  
 Given (\emph{Non-GMW-Entangled of Systems}), for system $x$ there is some $r\in\{1,2,\ldots, N\}$ and some degree-$r$ projector $P$ such that dim$(P)=r$ and $E(x, \sigma^r_r(P))$.  By (\emph{Eigenvector-Eigenvalue Link}), we can therefore attribute to $x$ the state $|\phi_1\rangle\wedge|\phi_2\rangle\wedge\ldots\wedge|\phi_r\rangle$, where span$(\{|\phi_i\rangle\ |\ i\in\{1,2,\ldots, r\}\}) = $  ran$(P)$.  This state is an eigenstate only of projectors of degree-$r$; so by (\emph{Eigenvector-Eigenvalue Link}) again, if $E(x,Q)$ for any projector $Q$, then $Q$ has degree $r$.
 $\Box$

\begin{definition}[Degree of Systems]\label{RankS}
For any system $x$, $\deg(x)$ is the unique degree of any projector $P$ such that $E(x, P)$.
\end{definition}

\begin{proposition}[Reflexivity of $\sqsubseteq$]\label{reflex}
For any system $x$, $x\sqsubseteq x$.
\end{proposition}
\emph{Proof.}
This follows straightforwardly from (\emph{Definition of Parthood}).
$\Box$

\begin{proposition}[Transitivity of $\sqsubseteq$]\label{trans}
For any systems $x,y, z$, if $x\sqsubseteq y$ and $y\sqsubseteq z$, then $x\sqsubseteq z$.
\end{proposition}
\emph{Proof.}
This follows straightforwardly from (\emph{Definition of Parthood}).
$\Box$

\begin{proposition}[State-System Uniqueness 1]\label{StateU1}
For any degree-1 projector $P$, if there is some system $x$ such that $E(x,\sigma^r_r(P))$, where $r = $ dim$(P)$, then $x$ is unique.
\end{proposition}
\emph{Proof.}
Let $P$ be any degree-1 projector with dim$(P)=r$. Assume that there is an $x$ such that $E(x,\sigma^r_r(P))$.  By (\emph{Existence of Subsystems}), $E(\Omega, \sigma^N_r(P))$.  Since dim$(P) = r$, $\sum_{i=r}^N \sigma^N_i(P) = \sigma^N_r(P)$; so $E(\Omega, \sum_{i=r}^N \sigma^N_i(P))$.  By (\emph{Uniqueness of Subsystems}), there is a unique system $y$ such that $E(y,\sigma^r_r(P))$; so $x=y$ and $x$ is unique.
$\Box$

\begin{proposition}[State-System Uniqueness 2]\label{StateU2}
For any system $x$, the degree-1 projector $P$ such that $E(x,\sigma^r_r(P))$, where $r = $ dim$(P)$, is unique.
\end{proposition}
\emph{Proof.}
Let $x$ be any system.  By (\emph{Non-GMW-Entangled Systems}), there is some degree-1 projector $P$ such that $E(x,\sigma^r_r(P))$, where $r = $ dim$(P)$. Suppose for \emph{reductio} that there is some other degree-1 projector $Q$ with dim$(Q) = r$ such that $E(x,\sigma^r_r(Q))$.  $Q\nleqslant P$ and $P\nleqslant Q$, since $P\neq Q$ and dim$(P) = $ dim$(Q)$.  So by (\emph{Definition of Parthood}), $x\nsqsubseteq x$, which contradicts Proposition \ref{reflex} (\emph{Reflexivity of $\sqsubseteq$})
$\Box$

This allows us to attribute to each system a pure state, as follows:
\begin{definition}[Subsystem States]\label{StateDef}
For any system $x$, the state of $x$ is the unique projector $\sigma^r_r(P)$ such that $r = \dim(P) = \deg(x)$ and $E(x,\sigma^r_r(P))$.
\end{definition}
With this definition we can extend application of (\emph{Eigenstate-Eigenvalue Link}) to systems other than $\Omega$.

\begin{proposition}[Anti-Symmetry of $\sqsubseteq$]\label{antisym}
For any two systems $x,y$, if $x\sqsubseteq y$ and $y\sqsubseteq x$, then $x=y$.
 \end{proposition}
\emph{Proof.}
 (\emph{Definition of Parthood}) and Proposition \ref{RankU} (\emph{Unique Degree}) entail: if $x\sqsubseteq y$ and $y\sqsubseteq x$, then for
all degree-1 projectors $P$ and all $r\in\{1,2,\ldots, N\}$, $E(x,\sigma^r_r(P))$ iff $E(y, \sigma^r_r(P))$.  From (\emph{Non-GMW-Entangled  Systems}), there is some degree-1 projector $Q$ such that $E(x,\sigma^r_r(Q))$ and dim$(Q) = r$.  So also $E(y,\sigma^r_r(Q))$.  By Proposition \ref{StateU1} (\emph{State-System Uniqueness 1}), $x=y$.
$\Box$

Propositions \ref{reflex}, \ref{trans} and \ref{antisym} entail that parthood for fermions is a partial ordering relation, thereby satisfying the first mereological axiom.

\begin{corollary}[Criterion of Identity]\label{CoI}
For any systems $x, y$, $x=y$ iff: for all projectors $P$ and all $r\in\{1,2,\ldots, N\}$, $E(x, \sigma^r_r(P))$ iff $E(y, \sigma^r_r(P))$.
\end{corollary}
In the particular case in which $\deg(x)$ = $\deg(y)$ = 1, this entails a satisfyingly straightforward statement of the Pauli Exclusion Principle: $x=y$ iff, for all degree-1 projectors $P$,  $E(x,P)$ iff $E(y,P)$.  Another interesting consequence is that  in any non-GMW-entangled joint state, any two individual fermions are  discernible by monadic predicates (which Muller \& Saunders (2008) call \emph{absolutely discernible}).  This is contrary to the orthodoxy in the quantum literature, in which bosons and fermions are taken to be \emph{either merely weakly discernible or utterly indiscernible}  (French \& Redhead 1988; Butterfield 1993; Huggett  2003; French \& Krause 2006; Muller \& Saunders 2008; Muller \& Seevinck 2009; Caulton 2013). But this orthodoxy relies on adopting an alternative interpretation of permutation invariance, in which we can still give physical meaning to the labels of the factor Hilbert spaces in the joint Hilbert space.  On our strict reading of permutation-invariance,  fermions may be  individuated by their states, as suggested by Dieks \& Lubberdink (2011).

\begin{proposition}[$\Omega$ is Maximal]\label{Omega}
Every system is a part of the total system $\Omega$. 
\end{proposition}
\emph{Proof.} Take $\Omega$.  Its state is $|\psi_1\rangle\wedge|\psi_2\rangle\wedge\ldots\wedge|\psi_N\rangle$.  By (\emph{Eigenstate-Eigenvalue Link}), it follows that $E(\Omega, \sigma^N_N(P))$, where ran$(P)$ = span$(\{|\psi_i\rangle\})$.  Since dim$(\sigma^N_N(P)) = 1$ and because of (\emph{Eigenstate-Eigenvalue Link}), any other degree-1 projector $Q$ such that $E(\Omega, \sigma^N_N(Q))$ must satisfy $\sigma^N_N(P))\leqslant  \sigma^N_N(Q)$, and so $P\leqslant Q$.

Now take any system $x$. From (\emph{Non-GMW-Entangled Systems}), there is some $r\in\{1,2,\ldots, N\}$ and some degree-1 projector $R$ such that $E(x, \sigma^r_r(R))$ and dim$(R) = r$. By Proposition \ref{StateU1} (\emph{State-System Uniqueness 1}), $x$ is unique.  We may now use (\emph{Uniqueness of Subsystems}) to infer $E(\Omega, \sigma^N_r(R))$.  But from the previous paragraph, we must have $\sigma^N_N(P) \leqslant \sigma^N_r(R)$.  So, by Proposition \ref{ProjResult}, $R\leqslant P$.  From this and (\emph{Eigenstate-Eigenvalue Link}), it follows that $E(x, \sigma^r_r(P))$.  And by  (\emph{Eigenstate-Eigenvalue Link}) again, for any degree-1 projector $Q$ such that $P\leqslant Q$, $E(x, \sigma^r_r(Q))$.

From the two preceding paragraphs it follows that, for any degree-1 projector $Q$ such that $E(\Omega, \sigma^N_N(Q))$, we also have $E(x, \sigma^r_r(Q))$.  So, by Proposition \ref{RankU} (\emph{Unique Rank}) and (\emph{Definition of Parthood}), $x\sqsubseteq\Omega$. $\Box$

\begin{definition}[System-Spaces]\label{SpaceS}
For any system $x$, the system-space $\mathfrak{s}(x)$ associated with $x$ is the range of the unique degree-1 projector $P$ such that $\dim(P) = \deg(x) = : r$ and $E(x, \sigma^r_r(P))$.
\end{definition}
So, as anticipated in Section \ref{GMW}, any system $x$ is associated with a  subspace of the single-particle Hilbert space $\mathcal{H}$.  Moreover, for any system $x$, dim$(\mathfrak{s}(x)) = \deg(x)$.  In general, for any system $x$, we may write $x$'s state as $|\phi_1\rangle\wedge|\phi_2\rangle\wedge\ldots\wedge |\phi_r\rangle$, where $r := \deg(x)$ and the $\{|\phi_i\rangle\}$ are orthonormal; then $\mathfrak{s}(x)$ = ran$\left(\sum_{i=1}^r|\phi_i\rangle\langle\phi_i|\right)$.

\begin{proposition}[Subspacehood Represents  Parthood]\label{SP}
For any systems $x,y$, $\mathfrak{s}(x) \subseteq \mathfrak{s}(y)$ iff $x\sqsubseteq y$. 
\end{proposition}
\emph{Proof.}
\emph{Left to Right:} Assume $\mathfrak{s}(x) \subseteq \mathfrak{s}(y)$. Given Proposition \ref{StateU2} (\emph{State-System Uniqueness 2}), let $P$ be the unique degree-1 projector such that $\dim(P) = \deg(x)=:r$ and $E(x, \sigma^r_r(P))$ and $Q$ be the unique degree-1 projector such that $\dim(Q) = \deg(y)=:s$ and $E(y, \sigma^s_s(Q))$.  Since $\mathfrak{s}(x) \subseteq \mathfrak{s}(y)$, $P\leqslant Q$ and $r\leqslant s$.   $\dim(\sigma^s_s(Q)) = 1$, so any  degree-1 projector $R$ such that $E(y, \sigma^s_s(R))$ must be such that $Q\leqslant R$, whence $E(x,\sigma^r_r(R))$.  From (\emph{Definition of Parthood}), it follows that $x\sqsubseteq y$.

\emph{Right to Left:}  Assume $x\sqsubseteq y$.  Given Proposition \ref{StateU2} (\emph{State-System Uniqueness 2}), let $P$ be the unique degree-1 projector such that $\dim(P) = \deg(x)=:r$ and $E(x, \sigma^r_r(P))$ and $Q$ be the unique degree-1 projector such that $\dim(Q) = \deg(y)=:s$ and $E(y, \sigma^s_s(Q))$.   From (\emph{Definition of Parthood}) and Proposition \ref{RankU} (\emph{Unique Degree}), it follows that $E(x, \sigma^r_r(Q))$.  But $\dim( \sigma^r_r(P)) = 1$, so $P\leqslant Q$; whence $\mathfrak{s}(x) \subseteq \mathfrak{s}(y)$.
$\Box$

\begin{corollary}[System-Subspace Link]\label{SSLink}
For any systems $x,y$, $\mathfrak{s}(x) = \mathfrak{s}(y)$ iff $x=y$. 
\end{corollary}
\emph{Proof.}
The Right to Left direction is trivial. 
\emph{Left to Right:} Assume $\mathfrak{s}(x) = \mathfrak{s}(y)$. Then $\mathfrak{s}(x) \subseteq \mathfrak{s}(y)$ and $\mathfrak{s}(y) \subseteq \mathfrak{s}(x)$. It follows from Proposition \ref{SP} (\emph{Subspacehood Represents  Parthood}) and Proposition \ref{antisym} (\emph{Anti-Symmetry of $\sqsubseteq_f$})  that $x=y$.
$\Box$

\begin{proposition}[Each System-Space is a Subspace of $\mathfrak{s}(\Omega)$]\label{SS1}
For any system $x$, $\mathfrak{s}(x)\subseteq \mathfrak{s}(\Omega)$. 
\end{proposition}
 \emph{Proof.} This follows straightforwardly from Proposition \ref{Omega} (\emph{$\Omega$ is Maximal}) and Proposition \ref{SP} (\emph{Subspace Represents  Parthood}). $\Box$

\begin{proposition}[Each Subspace of $\mathfrak{s}(\Omega)$ is a System-Space]\label{SS2}
For any non-zero space $\mathfrak{x} \subseteq \mathfrak{s}(\Omega)$, there is a unique $x$ such that $\mathfrak{s}(x) = \mathfrak{x}$. 
\end{proposition}
 \emph{Proof.} 
 Take any non-zero space $\mathfrak{x} \subseteq \mathfrak{s}(\Omega)$.  This defines the degree-1 projector $P$ for which ran$(P) = \mathfrak{x}$.  Let $r:=$ dim$(P)$.  $\mathfrak{s}(\Omega)$ similarly defines the degree-1 projector $Q$ for which ran$(Q) = \mathfrak{s}(\Omega)$ and dim$(Q) = N$.  We  know from Definition \ref{SpaceS} (\emph{System-Spaces}) that $E(\Omega, \sigma^N_N(Q))$.  And, since  $\mathfrak{x} \subseteq \mathfrak{s}(\Omega)$, $P\leqslant Q$;  and so, by Proposition \ref{ProjResult}, $\sigma^N_N(Q)\leqslant \sigma^N_r(P)$. Using (\emph{Eigenstate-Eigenvalue Link}) we may infer that $E(\Omega, \sigma^N_r(P))$.  From (\emph{Uniqueness of Subsystems}) it follows that there is a unique system $x$ such that $E(x,\sigma^r_r(P))$.  From Definition \ref{SpaceS} (\emph{System-Spaces}) it follows that $\mathfrak{s}(x) = \mathfrak{x}$.
 $\Box$
 
 The foregoing results show that, given our non-GMW-entangled $N$-fermion assembly $\Omega$, the totality of all non-GMW-entangled systems in existence correspond one-to-one to the subspaces of $\mathfrak{s}(\Omega)$, i.e.~the elements of the exterior algebra $\Lambda(\mathfrak{s}(\Omega))$ (except for the zero subspace).  Using the fact that parthood is represented by subspacehood, we can infer the representations of the other mereological notions: overlap, product.  Two objects overlap iff they possess a common part; so two systems overlap iff their systems spaces have a non-zero intersection.  The mereological product $x\sqcap y$ of any two systems $x$ and $y$ (if it exists), is then a system whose system-space is the intersection of the two corresponding system-spaces.  this greatly clarifies the compositional structure of non-GMW-entangled fermionic states.

\begin{proposition}[Parthood Obeys Atomicity]
Every system  has some part that has no proper parts.
\end{proposition}
\emph{Proof.}
We use Propositions \ref{SP} (\emph{Subspacehood Represents Parthood}), \ref{SS1} (\emph{Each System-Space is a Subspace of $\mathfrak{s}(\Omega)$}) and \ref{SS2} (\emph{Each Subspace of $\mathfrak{s}(\Omega)$ is a System-Space}).  Take any system $x$.  $x$ has a system-space $\mathfrak{s}(x)$ which is a subspace of $\mathfrak{s}(\Omega)$.   $\mathfrak{s}(x)$ is spanned by $\deg(x)$-many 1-dimensional subspaces of  $\mathfrak{s}(\Omega)$; each one corresponds to a degree-1 system.  Since parthood is represented by subspacehood,  degree-1 systems have no proper parts.
$\Box$
 
 \begin{proposition}[Parthood Obeys Strong Supplementation]
For any systems $x$ and $y$, if  $x$ is not a part of $y$, then some part of $x$ is disjoint from $y$, i.e.~there is some system $z$ such that $z\sqsubseteq x$ and $\mathfrak{s}(z)\cap\mathfrak{s}(y) = \emptyset$.
\end{proposition}
\emph{Proof.}  Assume that $x\nsqsubseteq y$.  So by Proposition \ref{SP} (\emph{Subspacehood Represents Parthood}),  $\mathfrak{s}(x)\nsubseteq \mathfrak{s}(y)$.  Then there is some subspace $\mathfrak{z}$ of $\mathfrak{s}(x)$ such that $\mathfrak{z}\cap\mathfrak{s}(y) = \emptyset$.  By Proposition \ref{SS2} (\emph{Each Subspace of $\mathfrak{s}(\Omega)$ is a System-Space}), $\mathfrak{z} = \mathfrak{s}(z)$ for some system $z$.  By (\emph{Subspacehood Represents Parthood}) again, $z$ and $y$ are disjoint.
$\Box$

Thus we have proven all of our mereological axioms, except the axiom schema \emph{Unrestricted Fusion}.  The correspondence between systems and the elements of the exterior algebra $\Lambda(\mathfrak{s}(\Omega))$ leads to a surprising result:

\begin{proposition}[Continuum-Many Atomic Parts]
For any system $x$, if $\deg(x)\geqslant 2$, then $x$ has continuum-many atomic parts.
\end{proposition}
\emph{Proof.}
If  $\deg(x)\geqslant 2$, then $\dim(\mathfrak{s}(x))\geqslant 2$.  So there are continuum-many 1-dimensional subspaces $\mathfrak{y}\subset \mathfrak{s}(x)$.  Each one corresponds to a degree-1 (and therefore atomic) system. 
$\Box$

A vivid example is provided by the 2-fermion system $\Omega_2$ in the  spin-singlet state
\begin{equation}
|\!\uparrow\rangle\wedge|\!\downarrow\rangle = \frac{1}{\sqrt{2}}\left(|\!\uparrow\rangle\otimes |\!\downarrow\rangle - |\!\downarrow\rangle\otimes |\!\uparrow\rangle\right)
\end{equation}
It is well-known that this state is spherically symmetric, in that $|\!\uparrow\rangle\wedge|\!\downarrow\rangle = |\!\leftarrow\rangle\wedge|\!\rightarrow\rangle = \ldots$, for any pair of oppositely-pointing spin-$\frac{1}{2}$ states.  In our framework, this spherical symmetry reflects the fact that $\Omega_2$ possesses a continuum-multitude of atomic parts: one of which is spin-up, one is spin-down, one is spin-left, etc.

We are now ready to see the conflict with classical mereology.  For simplicity, I use the example of degree-1 systems.
\begin{proposition}[Non-Existence of Mereological Fusions 1]\label{FF1}
For any two degree-1 systems $x$ and $y$: if $\mathfrak{s}(x)\nsubseteq\mathfrak{s}(y)$, then there does not exist a system $z$ which is the mereological fusion $x\sqcup y$ of $x$ and $y$, i.e.~which is such that  $\forall w(w\circ z \leftrightarrow (w\circ x \vee w\circ y))$.
\end{proposition}
\emph{Proof.}
Since $x$ and $y$ are degree-1 systems and therefore atomic, $\mathfrak{s}(x)\nsubseteq\mathfrak{s}(y)$ entails that $x$ and $y$ are disjoint.  Since all systems' states are represented by subspaces, $z$'s state is represented by a subspace.  So we seek a subspace $\mathfrak{z}\subseteq\mathfrak{s}(\Omega)$ such that
\begin{equation}\label{FF1eq}
\mbox{For all } \mathfrak{w}\subseteq\mathfrak{s}(\Omega), \quad \mathfrak{w}\cap \mathfrak{z}\neq\emptyset \quad \mbox{iff}\quad (\mathfrak{w}\cap \mathfrak{s}(x)\neq\emptyset \quad\mbox{or}\quad \mathfrak{w}\cap \mathfrak{s}(y)\neq\emptyset ) .
\end{equation}
$\mathfrak{z}$ must have dimension at least 2, since it must overlap both $\mathfrak{s}(x)$ and $\mathfrak{s}(y)$.
But now consider some 1-dimensional subspace $\mathfrak{w}_0\subseteq\mathfrak{z}$ which is skew to (i.e.~neither coincident with nor orthogonal to)  both $\mathfrak{s}(x)$ and $\mathfrak{s}(y)$; such a subspace will always exist (take e.g.~the normalised sum proportional to $\mathfrak{s}(x) + \mathfrak{s}(y)$).  $\mathfrak{w}_0$ overlaps $\mathfrak{z}$, yet overlaps neither $\mathfrak{s}(x)$ nor  $\mathfrak{s}(y)$.   So no $\mathfrak{z}$ exists such that (\ref{FF1eq}) is satisfied.
$\Box$

\begin{proposition}[Non-Existence of Mereological Fusions 2]\label{FF2}
There are some satisfied 1-place formulas $\phi$ such that there is no system $x$ for which $\mathcal{F}_\phi(x)$.
\end{proposition}
\emph{Proof.}
Let $x$ and $y$ be any two distinct degree-1 systems, and let $P$ and $Q$ be the degree-1 projectors such that $\dim(P)=\dim(Q)=1$ and $E(x, P)$ and $E(y,Q)$.  Recall that, for any $\phi$, the fusion of the $\phi$s is defined by
\begin{equation}\label{Fdef}
\forall z(\mathcal{F}_\phi(z) \leftrightarrow \forall w(w\circ z \leftrightarrow \exists t(\phi(t)\ \&\ t \circ w)))
\end{equation}
 We  now set $\phi(t) := \left(E(t,P)\vee E(t,Q)\right)$.  $\phi(t)$ is satisfied by $x$ and $y$ only:  for  all systems are  individuated by their states, given Corollary \ref{CoI} (\emph{Criterion of Identity}).  So $\phi(t)$ is equivalent to $(t= x \vee t= y)$. By (\ref{Fdef}), the  fusion $z$ of the $\phi$s satisfies $\forall w(w\circ z \leftrightarrow (w\circ x \vee w\circ y))$.  By Proposition \ref{FF1} (\emph{Non-Existence of Mereological Fusions 1}), no such $z$ exists.
$\Box$

The failure of \emph{Unrestricted Fusion} might seem surprising, since we  have a way of producing a fermionic joint state out of any collection of degree-1 fermion states: this is given by the wedge product, as discussed in Section \ref{GMW}.  What is going on here is that the corresponding notion of composition is \emph{not} mereological.  

Take any systems $x$ and $y$.  There are two degree-1 projectors $P$ and $Q$ such that  $\dim(P) = \deg(x) =:r$, $\dim(Q)=\deg(y)=: s$, $E(x,\sigma^r_r(P))$ and $E(y,\sigma^s_s(Q))$.  Now let $\Sigma(P,Q)$ be the degree-1 projector whose range is the span of the ranges of $P$ and $Q$.  Then we may define
\begin{definition}[Fermionic Fusion]\label{FF}
For any systems $x$ and $y$ and associated degree-1 projectors $P$ and $Q$, the fermionic fusion of $x$ and $y$, denoted $x +_f y$, is the unique system $z$ such that $E(z, \sigma^t_t(\Sigma(P,Q)))$, where $t = \dim\Sigma(P,Q)$.
\end{definition}
The existence and uniqueness of fermionic fusion is guaranteed by Propositions \ref{SS2} (\emph{Each Subspace of $\mathfrak{s}(\Omega)$ is a System-Space}) and \ref{StateU1} (\emph{State-System Uniqueness 1}).  This constitutes a fermionic analogue to \emph{Unrestricted Fusion}.

To better understand fermionic fusion, we note that
$x+_f x = x$, 
and, denoting the vector-state of any system $x$ by $\mathfrak{v}(x)$, 
\begin{proposition}[Wedge product and fermionic fusion]
If $x$ and $y$ are degree-1 systems such that $\mathfrak{v}(x)\perp\mathfrak{v}(y)$, then $\mathfrak{v}(x+_f y) = \mathfrak{v}(x)\wedge \mathfrak{v}(y)$.
\end{proposition}
\emph{Proof.}
Since $\mathfrak{v}(x)\perp\mathfrak{v}(y)$, $\mathfrak{v}(x)\wedge\mathfrak{v}(y)$ corresponds to a correctly normalised anti-symmetric vector, corresponding to the space spanned by $\mathfrak{v}(x)$ and $\mathfrak{v}(y)$. 
$\Box$

\begin{figure}[h!] 
\begin{center}
\setlength{\unitlength}{1mm}  
\begin{picture}(100,60)   
\put(20,0){\includegraphics[width=0.4\textwidth]{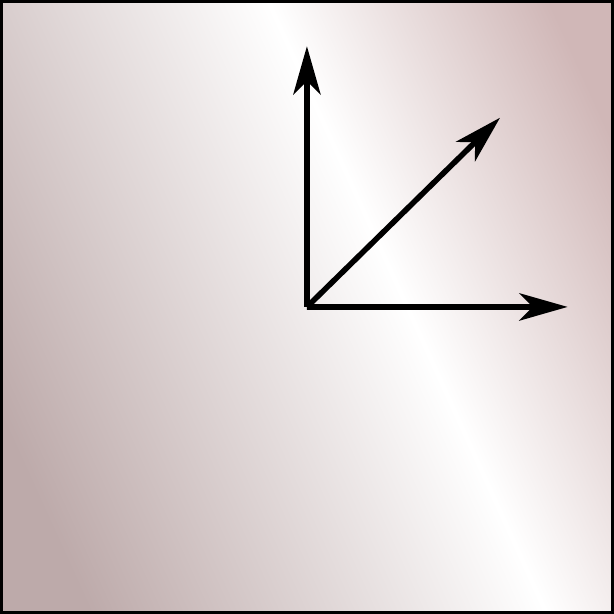}}
\put(81,2){$|\!\uparrow\rangle\wedge|\!\downarrow\rangle$}
\put(70,23){$|\!\uparrow\rangle$}
\put(42,50){$|\!\downarrow\rangle$}
\put(65,38){$|\!\rightarrow\rangle$}

\end{picture}
\end{center}
\caption{Three coplanar vectors, corresponding to degree-1 fermionic systems in the states  $|\!\uparrow\rangle$,  $|\!\downarrow\rangle$  and $|\!\rightarrow\rangle := \frac{1}{\sqrt{2}}(|\!\uparrow\rangle +|\!\downarrow\rangle)$.  The plane corresponds to a degree-2 fermionic system in the state $|\!\uparrow\rangle\wedge|\!\downarrow\rangle$, which is the fermionic fusion of any pair of the three degree-1 systems.  Mereological product does not distribute over fermionic fusion; in this example represented by the fact that the intersection of $|\!\rightarrow\rangle$ with the span of $|\!\uparrow\rangle$ and $|\!\downarrow\rangle$ ($= |\!\rightarrow\rangle$) is not equal to the span of the intersections of $|\!\rightarrow\rangle$ with $|\!\uparrow\rangle$ and $|\!\downarrow\rangle$ ($=\mathbf{0}$). \label{3vectors}} 
\end{figure}
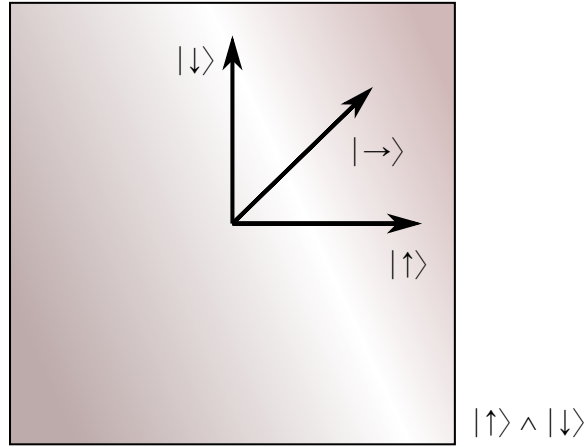

\begin{proposition}[Failure of Distributivity]
For not all  systems $x,y,z$: $x\sqcap (y+_f z) = (x\sqcap y) +_f (x\sqcap z)$.
\end{proposition}
\emph{Proof.}
It suffices to give an example of three degree-1 systems all of whose states are  coplanar; see Figure \ref{3vectors}.
$\Box$

At this point we see a strong analogy between the structure of fermionic composition and the quantum logic of Birkoff and von Neumann (1936).  For a fuller discussion of quantum logic (particularly subtleties involving infinite-dimensional Hilbert space), see Dalla Chiara, Giuntini \& R\'edei (2007).  Here it will suffice to draw the following analogies between relations between and operations on objects and propositions, both classical and quantum:

\begin{tabular}{c|c||c|c}
\emph{Classical composition} & \emph{Classical logic} & \emph{Fermionic composition} & \emph{Quantum logic} \\
\hline\hline&&&\\
Classical system & Classical  & Fermionic  & Quantum  \\
& proposition & system & proposition\\&&&\\
Total system & Tautology  & Total system $\Omega$ & Tautology\\&&&\\
Parthood $\sqsubseteq$ & Classical  & Parthood $\sqsubseteq$ & Quantum  \\
& entailment &&entailment\\&&&\\
Mereological&  Classical  & Mereological & Quantum \\
 product $\sqcap$ & conjunction $\wedge$&  product $\sqcap$ & conjunction\\&&&\\
Mereological&  Classical  & Fermionic & Quantum \\
 fusion $\sqcup$ & disjunction $\vee$&  fusion $+_f$ & disjunction\\&&&\\
 Classical & Classical & Fermionic & Quantum \\
 complement & negation $\neg$ & complement & negation (ortho-\\
 &&& complement) $\perp$
\end{tabular}

In each row, analogies between columns 1 and 2 and columns 3 and 4 respectively are exact insofar as they receive the same mathematical representation, in Boolean algebras and Hilbertian lattices, respectively.  Analogies between columns 1 and 3 or between columns 2 and 4 are looser.

A final comment.  For those of us who wish to think of the middle-sized dry goods of everyday life as ``made up'' of fermions, it might appear at first blush as something of a mystery how to reconcile the Hilbertian structure of fermionic composition with the Boolean structure of our heuristic understanding of the composition of middle-sized dry goods.  In fact there need be no mystery here: macroscopic objects are individuated (at least approximately) by their spatial boundaries.  This picks a preferred orthobasis in the Hilbertian lattice of fermionic states, and the subspaces spanned by the rays in this basis have the familiar structure of a Boolean algebra.

\subsection{Can mereology be saved?}\label{Saving}

The idea that quantum mechanics might prompt a revision in logic, a view argued by Putnam (1969, 1974), has received rather short shrift.  For example, here is
Jauch (1968), quoted in Dalla Chiara, Guintini and R\'edei:
\begin{quote}
The propositional calculus of a physics system has a certain similarity to the corresponding calculus of ordinary logic.  In the case of quantum mechanics, one often refers to this analogy and speak of quantum logic in contradistinction to ordinary logic. \ldots  The calculus introduced here has an entirely different meaning from the analogous calculus used in formal logic.  Our calculus us the formalization of a set of \emph{empirical} relations which are obtained by making measurements on a physical system.  It expresses an objectively given property of the physical world.  It is thus the formalization of empirical facts, inductively arrived at and subject to the uncertainty of any such fact.  The calculus of formal logic, on the other hand, is obtained by making an analysis of the meaning of propositions.  It is true under all circumstances and even tautologically so.  Thus, ordinary logic is used even in quantum mechanics of systems with a propositional calculus vastly different from that of formal logic.  The two need have nothing in common.
\end{quote}
One might wish to say the same of mereological and fermionic composition.  That is, although the mathematical theory associated with fermions suggests a particular calculus with similarities to---but crucial differences from---classical mereology, the option seems to be open simply to deny that anything other than mereological composition is composition worth the name.   Such a strategy will be friendly, if not downright essential, to anyone who takes classical mereology to be  `perfectly understood, unproblematic, and certain' (Lewis (1991, 75)).

In fact this strategy is possible, and could proceed by simply admitting the existence of the mereological fusions currently ruled out.  That is, we expand the domain to include not only the fermionic systems, but also  arbitrary  fusions thereof.  A typical such fusion will \emph{not} be a system in the sense that its state, if one can be attributed to it at all, will not be representable as a vector in  $\mathcal{A}(\otimes^r\mathcal{H})$, for any $r$.  

To get a better idea of these non-system objects, recall that, 
following Kibble (1979) and Ashtekar \& Schilling (1999), we may describe the possible states of a degree-1 quantum system not with the unit-vectors of the single-system Hilbert space $\mathcal{H}$, but rather by the points of the projective Hilbert space $\mathcal{P(H)}$.  Given the correspondence proven above between non-GMW-entangled $N$-fermion states and $N$-dimensional subspaces of $\mathcal{H}$, and the well-defined map between rays of $\mathcal{H}$ and points of $\mathcal{P(H)}$, we may carry over the above results to represent arbitrary non-GMW-entangled fermion states as \emph{regions} of $\mathcal{P(H)}$---indeed they will be regions that are also subspaces of $\mathcal{P(H)}$.

The (singletons of) points and subspaces of $\mathcal{P(H)}$ do not constitute a Boolean algebra: this is the mathematical expression of the failure of classical mereology.
But we can add more subsets of $\mathcal{P(H)}$ \emph{until} we achieve a Boolean algebra.  For example, for any two singletons $\{\psi\}, \{\phi\}$, where $\phi, \psi\in\mathcal{P(H)}$, we add their union $\{\psi, \phi\}$.  We associate this union with the mereological fusion of the degree-1 systems whose states are given by $\psi$  and $\phi$.
Given the fact that the mereological axioms we have been considering are first-order, we may recover the truth of \emph{Unrestricted Fusion} without having to admit the full power set of $\mathcal{P(H)}$; in fact arbitrary finite unions of  (singletons of) points and subspaces will do.

It is hardly any objection that these non-system objects are somehow unnatural or that we have no practical use for them in any scientific theory:  that is a familiar feature of arbitrary mereological fusions.  However, what is different in this case is that a \emph{rival} notion of composition, i.e.~fermionic composition, still runs alongside the classical one; and it is the fermionic concept that produces familiar states from familiar states.  The bizarre non-system objects may be admitted or they may not; there is just no getting around that fact that the compositional structure of the fermionic \emph{systems} is Hilbertian, rather than Boolean.

\section{Conclusion}\label{Conc}

The foregoing arguments can be summarised as follows: \emph{at least one} of the following three claims must be rejected:

\begin{enumerate}
\item Permutation invariance reflects representational redundancy.
\item Fusions of fermionic systems are always fermionic systems.
\item Fermions compose (i.e.~fuse) mereologically.
\end{enumerate}

Premise 1 was crucial to our new way of thinking of entanglement for fermionic systems.  By rejecting the strong reading of permutation-invariance, we may retrench to an identification of systems with factor Hilbert spaces.  That way we avoid the permutation-invariant method for identifying subsystems, and the resulting failure of mereology. But rejecting premise 1 is intolerable, because it means giving an interpretation to permutation invariance that does not best explain it.  Our best way of understanding permutation-invariance is precisely as reflecting a representational redundancy, or so I argued in Section \ref{PI}.

Premise 2 may be held by anyone who is sanguine about the possibility that our best theory of composition might be informed by empirical science.  By rejecting it, we may take the saving strategy discussed in Section \ref{Saving}, and hang on to both our strong reading of permutation-invariance and our conviction that mereological composition is the only composition worth the name.  But rejecting it might seem intolerable, on pain of admitting new, strange objects into our ontology in whose existence  we have no independent reason to believe. 

We have seen that the natural mathematical structure of fermionic states poses a threat to premise 3. This threat is not compelling, insofar as premises 1 and 2 are not compelling.  And rejecting premise 3 will be intolerable to anyone who takes mereology to be `perfectly understood, unproblematic, and certain'.  The question of whether our traditional understanding of composition is immune to the deliverances of quantum mechanics   hangs on which intolerable claim one is prepared to accept.

\section{References}
\ \\
\parindent=-12pt
Ashtekar, A., and  Schilling, T. A., (1999), `Geometrical formulation of quantum mechanics' in \emph{On Einstein's Path,} New York: Springer, pp.~23-65.

Bacciagaluppi, G. (2009), `Is Logic Empirical?', in K. Engesser, D. Gabbay \& D. Lehmann (eds), \emph{Handbook of Quantum Logic and Quantum Structures: Quantum Logic} (Elsevier, Amsterdam), pp. 49-78.

Birkhoff, G. \& von Neumann, J. (1936), `The logic of quantum mechanics', \emph{Annals of Mathematics} \textbf{37}, pp. 823-843.

Butterfield, J.~N.~(1993), `Interpretation and identity in quantum theory', \emph{Studies in the History and Philosophy of Science} \textbf{24}, pp. 443-76.

Caulton, A. (2013). `Discerning ``indistinguishable'' quantum systems', \emph{Philosophy of Science}
\textbf{80}, pp.~49-72.

Caulton, A. (2015), `Individuation and Entanglement in Permutation-Invariant Quantum Mechanics'.

Dalla Chiara, M. L.,  Giuntini, R., \&  R\'edei, M. (2007), `The History of Quantum Logic', 
in D. Gabbay \& J. Woods (eds.), \emph{Handbook of History of Logic, Vol. 8: The Many Valued and Nonmonotonic Turn in Logic} (North Holland), pp. 205-283.

Dummett, M. (1976), `Is Logic Empirical?', in H. D. Lewis (ed.), \emph{Contemporary British Philosophy}, 4th series (London: Allen and Unwin), pp. 45?68. Reprinted in M. Dummett, \emph{Truth and other Enigmas} (London: Duckworth, 1978), pp. 269-289.

French, S. and Krause, D. (2006), \emph{Identity in Physics: A Historical, Philosophical and Formal Analysis.}  Oxford: Oxford University Press.

French, S. and Redhead, M. (1988), `Quantum physics and the identity of indiscernibles', \emph{British Journal for the Philosophy of Science} \textbf{39}, pp.~233-46.

Ghirardi, G. and Marinatto, L. (2003), `Entanglement and Properties', \emph{Fortschritte der Physik}
\textbf{51}, pp.~379Ð387.

Ghirardi, G. and Marinatto, L. (2004), `General Criterion for the Entanglement of Two Indistinguishable Particles', \emph{Physical Review A} \textbf{70}, 012109-1-10.

Ghirardi, G. and Marinatto, L. (2005), `Identical Particles and Entanglement', \emph{Optics and Spectroscopy} \textbf{99}, pp.~386-390.

Ghirardi, G., Marinatto, L. and Weber, T. (2002),  `Entanglement and Properties of Composite Quantum Systems: A Conceptual and Mathematical Analysis', \emph{Journal of Statistical Physics} \textbf{108}, pp.~49-122.

Hovda, P. (2009),  `What Is Classical Mereology?', \emph{Journal of Philosophical Logic} \textbf{38}, pp. 55-82.

Huggett, N. (1999), `On the significance of permutation symmetry', \emph{British Journal for the Philosophy of Science} \textbf{50}, pp.~325-47.

Huggett, N. (2003), `Quarticles and the Identity of Indiscernibles', in K. Brading and E. Castellani (eds.), \emph{Symmetries in Physics: New Reflections}, Cambridge: Cambridge University Press, pp.~239-249.

Jauch, J. M. (1968), \emph{Foundations of Quantum Mechanics.} Addison-Wesley, London.

Kibble, T. W. B., (1979), `Geometrization of Quantum Mechanics', \emph{Communications in Mathematical Physics} \textbf{65}, pp.~189-201. 

Kochen, S. and Specker, E. P. (1967), `The Problem of Hidden Variables in Quantum Mechanics', \emph{Journal of Mathematics and Mechanics} \textbf{17}, pp.~59-87.

Ladyman, J., Linnebo, \O., and Bigaj, T. (2013), `Entanglement and non-factorizability', \emph{Studies in History and Philosophy of Modern Physics} \textbf{44}, pp.~215-221.

Leonard, H. S. and Goodman, N. (1940), `The Calculus of Individuals and Its Uses', \emph{Journal of Symbolic Logic} \textbf{5}, pp.~45-55.

L\'esniewski, S. (1916/1992), \emph{Podstawy og\'olnej teoryi mnogosci I.} Moskow: Prace Polskiego Kola Naukowego w Moskwie, Sekcya matematyczno-przyrodnicza. Trans. by D. I. Barnett (1992), `Foundations of the General Theory of Sets', in S. L\'esniewski, \emph{Collected Works, Vol. 1,}, ed. S. J. Surma, J. Srzednicki, D. I. Barnett, and F. V. Rickey, Dordrecht: Kluwer,  pp. 129-173.

Lewis, D. (1968), `Counterpart Theory and Quantified Modal Logic', \emph{Journal of Philosophy} \textbf{65}, pp.~113-126.

Lewis, D. (1991), \emph{Parts of Classes}.  Wiley-Blackwell.

Mac Lane, S. \& Birkoff, G. (1991), \emph{Algebra}, Third Edition.  AMS Chelsea.

Margenau, H. (1944), `The Exclusion Principle and its Philosophical Importance', 
\emph{Philosophy of Science} \textbf{11}, pp.~187-208. 

Maudlin, T. (1998), `Part and Whole in Quantum Mechanics', in E. Castellani (ed.), \emph{Interpreting Bodies: Classical and Quantum Objects in Modern Physics.} Princeton: Princeton University Press.

Maudlin, T. (2005), `The Tale of Quantum Logic', in Y. Ben-Menahem (ed.), \emph{Hilary Putnam: Contemporary Philosophy in Focus} (Cambridge University Press), pp.~156-187.

Messiah, A. M. L. and Greenberg, O. W. (1964), `Symmetrization Postulate and Its Experimental Foundation', \emph{Physical Review} \textbf{136}, pp.~B248-B267.

Muller, F.~A. and Saunders, S. (2008), `Discerning Fermions', \emph{British Journal for the Philosophy of Science}, \textbf{59}, pp.~499-548.

Muller, F. A. and Seevinck, M. (2009), `Discerning Elementary Particles', \emph{Philosophy of Science}, \textbf{76}, pp.~179-200.

Nielsen, M. A. and Chuang, I. L. (2010), \emph{Quantum Computation and Quantum Information}, 10th anniversary edition.  Cambridge: Cambridge University Press.

Putnam, H. (1968), `Is Logic Empirical?', in R. Cohen and M. Wartofsky (eds.), \emph{Boston Studies in the Philosophy of Science}, vol. 5 (Dordrecht: Reidel), pp. 216-241. Reprinted as `The Logic of Quantum Mechanics', in H. Putnam, \emph{Mathematics, Matter, and Method. Philosophical Papers, vol. 1} (Cambridge: Cambridge University Press, 1975), pp. 174-197.

Putnam, H. (1974), `How to Think Quantum-Logically', \emph{Synthese} \textbf{29}, pp.~55-61. Reprinted in P. Suppes (ed.), \emph{Logic and Probability in Quantum Mechanics} (Dordrecht: Reidel, 1976) pp. 47-53.

Russell, B. (1918/1985), `The Philosophy of Logical Atomism'.   Open Court.

Tarski, A. (1929), `Les fondements de la g\'eom\'etrie des corps', \emph{Ksiega Pamiatkowa Pierwszkego Polskiego Zjazdu Matematycznego}, supplement to \emph{Annales de la Soci\'et\'e Polonaise de Math\'ematique} \textbf{7}, pp. 29-33. Translated by J. H. Woodger (1956) `Foundations of the Geometry of Solids', in A. Tarski, \emph{Logics, Semantics, Metamathematics. Papers from 1923 to 1938}, Oxford: Clarendon Press, 1956, pp. 24-29.

Tung, W.-K. (1985), \emph{Group Theory in Physics.}  River Edge: World Scientific.

Varzi, A. (2014), `Mereology', \emph{The Stanford Encyclopedia of Philosophy (Spring 2014 Edition)}, Edward N. Zalta (ed.),\\ URL = $\langle$http://plato.stanford.edu/archives/spr2014/entries/mereology/$\rangle$.

\end{document}